\documentstyle[11pt]{article}
\begin{document}

\title{
\begin{flushright}
{\rm {\normalsize{USTC-ICTS-03-10}}}\\
\vspace{1cm}
\end{flushright}
BIon Configuration with Fuzzy $S^2$ Structure}

\author{ Yi-Xin Chen$^1$ \thanks{Email:yxchen@zimp.zju.edu.cn}
\hspace{.5mm} and Jing Shao$^{1,2}$ \\[.3cm]
$^1${\small Zhejiang Institute of Modern Physics, Zhejiang University,}\\
             {\small Hangzhou 310027, P.R.China,}\\
$^2${\small Interdisciplinary Center for Theoretical Study, University of Science and Technology of China,}\\
{\small Hefei, Anhui 230026, P.R.China} }

\date{}
\maketitle

\begin{abstract}

\indent

Some configurations of general dielectric D2-brane with both
electric and magnetic fields are investigated. We find two
classical stable solutions describing a BIon configuration with
$S^2$ structure and a cylindrical tube respectively. Both of them
can be regarded as the blown-up objects of the Born-Infeld string.
The dependence of the geometry of these configurations on the
Born-Infeld fields is analyzed. Also we find that similar BIon
configuration exists as a fuzzy object in a system with N
D0-branes. Furthermore this configuration is demonstrated to be
more stable than the usual dielectric spherical D2-brane.

\vspace{.5cm}

\end{abstract}

\setcounter{equation}{0}

\section{Introduction}

\indent

It is well known that the concept of D-brane \cite{Polchinski}
arises as the extended objects that open strings can end on. The
worldvolume dynamics of D-brane, which can be derived as the low
energy effective action of open string, is described by the
Born-Infeld (BI) action \cite{Leigh}. Dynamically, under some
circumstances, a lower-dimensional brane can be blown-up to
another higher-dimensional brane. Inversely, the lower-dimensional
brane can be considered as a partially collapsed version of the
higher-dimensional brane. It is observed in \cite{Emparan} that a
BI string can tunnel to a tubular D2-brane or nucleate to
spheroidal bulge of D2-brane in constant RR field background.
Another observation in \cite{Myers1} is that a similar non-trivial
background can blow up a bunch of D0-brane into fuzzy 2-sphere,
which is the well-known Myers effect. Furthermore,
\cite{Townsend1} suggests that string carrying D0-brane charge can
be 'blown-up' into supertube supported against collapse by
electric-magnetic angular momentum. The supertubes are collections
of type IIA fundamental strings and D0-branes, which have been
expanded to tubular 1/4-supersymmetric D2-branes by the addition
of angular momentum.

Born-Infeld theory admits some finite energy static solutions.
These solutions, called as BIons \cite{Gibbons}, are with
pointlike sources like the fundamental strings. They play the
important role in the theory of D-branes as the ends of string
intersecting the brane when the effects of gravity are ignored
\cite{Gibbons,Maldacena}. In particular, the BIon solution can be
used to analyze the tunnelling of some D-brane configurations.
Emparan \cite{Emparan} investigated some static unstable
configurations in the presence of a constant 4-form RR field
strength. These configurations include the bound configuration of
a spherical brane and the BI string approaching the limit of a
spike at large distance. This BI string possesses the character of
BIon solution. According to the BIon description of solutions in
Born-Infeld theory, we shall call the brane configuration with the
BIon's character as the BIon configuration. In this viewpoint, the
BIon configuration considered by Emparan is the bound
configuration of a D2-brane and the fundamental strings which can
be expressed as the sources of a Coulomb-like electric field. The
BIon solution in \cite{Townsend1} describes the configuration of
the type IIA strings ending on the D2-brane and D0-branes. In
order to preserve the supersymmetry, the 'dissolved' fundamental
strings which can be interpreted as the sources of the electric
field and the 'dissolved' D0-branes which are realized as the
magnetic fluxes must be considered. So this BIon configuration is
the bound brane configuration of a D2-brane, the fundamentle
strings and the D0-branes. Recently, Hyakutake \cite{Hyakutake1}
found the fuzzy BIon solution by establishing the dual description
of the BIon solution in \cite{Townsend1}. The fuzzy BIon is
described by the configuration obtained by pulling the fuzzy plane
in a fixed direction. It is natural to ask whether there exist
more general BIon solutions or not, and how the geometries of the
brane configurations are affected by the BIon's properties. In
this paper, we are interested in finding some new BIon solutions,
and analyzing the geometric characteristics of the BIon
configurations corresponding to them.

If we want to investigate more general BIon configurations, which
include the spherical D2-brane, the effective BI action considered
by us should include the Chern-Simons coupling term in the
presence of non-zero RR field strength. The reason is due to the
Myers effect \cite{Myers1} that the above non-trivial background
can blow-up a collection of D0-branes into a fuzzy two-sphere. In
the presence of the constant RR field strength, the matrix fields
of N coincident D0-branes' configuration are no longer globally
stable if they were diagonalized simultaneously. Also it is known
that in the supertube of \cite{Townsend1} the addition of
D0-branes, of which the charge can be interpreted as the magnetic
flux, is to provide a stabilization mechanism of preventing from
collapse by crossed electric and magnetic fields. In order to make
our considering BIon configuration stable, it should be the bound
brane configurations of a D2-brane, fundamental strings and
D0-branes in the background of the non-trivial RR field. Since the
fundamental strings can be expressed as the sources of the
electric field and the D0-brane charge can be interpreted as the
magnetic flux, our starting point should be the effective BI
action with both the electric field and the magnetic field in the
presence of the constant RR field strength.

In the next section, we shall discuss some BIon configurations
from the effective BI action as mentioned above. Indeed, We find
two classical stable solutions describing the BIon configurations
with $S^2$ structure and with tube structure respectively. The
former, which can be regarded as the bound brane configuration of
a spherical D2-brane, fundamental strings and D0-brane, is one of
the main results in this paper. Subsequently, in
terms of the dielectric effect from the viewpoint of $N$ coincident
D0-branes, we find that the BIon configuration with $S^2$
structure also exist as a fuzzy object. The energy of the BIon
configuration with the fuzzy $S^2$ structure is lower than that of
the configuration of the usual fuzzy $S^2$ found by Myers
\cite{Myers1}. That is, the former is more stable that the latter.

\section{BIon on a dielectric spherical D2-brane}

\indent

As mentioned above, here we focus on a D2-brane with N magnetic
fluxes in the presence of constant RR flux $F_{0123}^{4}=-4h$. For
generality we also allow the electric field on the world volume,
which can lead to the growing of a spike on the usual dielectric
spherical D2-brane. The BI action including the Chern-Simons
terms, which governs the low energy dynamics of D2-brane in a RR
background, reads as
\begin{equation}
I=-{T_2}\int d^3\xi \bigg\{ \sqrt{-\det(P[G]_{ab} + \lambda
F_{ab})} \bigg\}+\mu_2\int P[C^{(3)}],
\end{equation}
where $T_2=2\pi/(2\pi\l_s)^3 g_s$ is the D2-brane tension and
$\lambda=2\pi\l^2_s$. As usual $P[\cdots]$ is used to clarify the
the pullbacks, and the potential $C^{(3)}$ of the R-R field is
defined by $F^4 =dC^{(3)}$. Since we are interested in static
configuration with axial symmetry, it is convenient to use the
cylindrical coordinates
\begin{equation}
  x^0 = t, \quad x^1 = R(z) \cos\theta, \quad x^2 = R(z) \sin\theta, \quad
  x^3 = z,
\end{equation}
where $(t,\theta,z)$ are the world volume coordinates. For
time-independent magnetic field and electric field in z direction,
the BI 2-form field strength can be expressed as
\begin{equation}\label{Fieldstr1}
F=Edt \wedge dz+Bdz \wedge d\theta.
\end{equation}
Under the above conditions the D2-brane action can be reduced to
\begin{equation}
I=-{T_2}\int dt d\theta dz \bigg\{
\sqrt{R^2(-\lambda^2E^2+1)+\lambda^2B^2+R^2R^2_z}-2hR^2 \bigg\}.
\end{equation}

The Hamiltonian of the system is defined as
\begin{equation}
{\cal H} \equiv {\cal D}E-{\cal L}.
\end{equation}
Here ${\cal D}$ is the 'electric displacement' which is defined as
${\cal D}=\frac{1}{T_2\lambda} \frac{\partial L}{\partial E}$, and
it obeys the Gauss law constraint $\partial_z{\cal D}=0$. $E$ can
be express by using ${\cal D}$ as
\begin{equation}
E={{\cal D}\over\lambda
R}\sqrt{R^2(1+R^2_z)+\lambda^2B^2\over{\cal D}^2+R^2}.
\end{equation}
So the Hamiltonian density becomes
\begin{equation}
{\cal H}=T_2\bigg\{\frac{1}{R}\sqrt{({\cal D}^2+R^2)
(R^2(1+R^2_z)+\lambda^2B^2)}-2hR^2\bigg\}.
\end{equation}
The Euler-Lagrange equation for $A_x$ and $A_\theta$ can produce
another constraint on the magnetic field. If we define $
\Pi=-\frac{1}{T_2\lambda} \frac{\partial L}{\partial B}$, the
constraint reads as $\partial_z \Pi=\partial_\theta \Pi=0$. So
$\Pi$ is a constant. Furthermore, we can establish the relation
between the magnetic field $B$ and $\Pi$ in the following form
\begin{equation}\label{B}
B=\frac{\Pi R^2}{\lambda}\sqrt{1+R_z^2\over{\cal
D}^2+(1-\Pi^2)R^2}.
\end{equation}
It should be noticed from (\ref{B}) that the number of the
magnetic flux per unit area becomes infinitely small as $R$
approach to zero, unlike the case with no electric field
where it is uniform. Quantization of the magnetic flux requires
\begin{equation}\label{magflux}
N=\frac{1}{2\pi}\int d\theta dz B=\frac{1}{2\pi\lambda}\int
d\theta dz \sqrt{\frac{\Pi^2R^4(1+R_z^2)}{{\cal D}
^2+R^2-\Pi^2R^2}}.
\end{equation}
The Hamiltonian density can be rewritten in terms of $\Pi$ as
\begin{equation}\label{energy1}
{\cal H}=T_2\bigg\{({\cal D}^2+R^2)\sqrt{\frac{1+R_z^2} {{\cal
D}^2+(1-\Pi^2)R^2}}-2hR^2\bigg\}.
\end{equation}
Since the negative $h$ will oppose the expansion of the brane, we
will assume $h\geq 0$ in the following.

It is instructive to first consider the tube-like solution with
$R_z=0$ for all $z$. Since the configuration is noncompact, we
should assume that the length in the $z$ direction is taken as $L$
to make the energy finite. Using the constraint (\ref{magflux}) we
can eliminate $\Pi$ in the (\ref{energy1}) to give
\begin{equation}
{\cal H}=T_2\bigg\{\frac{1}{R}\sqrt{(R^2+\Omega^2)(R^2+{\cal
D}^2)}-2hR^2\bigg\},
\end{equation}
where $\Omega={\lambda N \over L}$. It is easily seen that for
nonzero $\Omega$ and ${\cal D}$, the BI-string has infinite
energy, thus is unstable. It would be blown-up to tube
configuration at local minimum of the energy. If $h=0$ this is
just the supertube solution. With $h$ turned on, the tube should
grow larger and become meta-stable, which can be understood as the
Myers effect. And when $h$ reach some critical value, the tube
would be completely unstable.

Now let us turn to general case with nonzero $R_z$ and look for
static solution by extremizing the energy with respect to the
field R. The static equation for R, which is gotten by varying
(\ref{energy1}) while N is fixed, is
\begin{equation}
\frac{\partial{\cal H}}{\partial R}+\frac{\partial{\cal
H}}{\partial \Pi}\frac{\partial\Pi}{\partial R}
-\frac{d}{dz}\frac{\partial{\cal H}}{\partial R_z}=0.
\end{equation}
After some algebraic manipulation, it can be rewritten as (only
valid for $R_z\neq 0$)
\begin{equation}
\frac{d}{dz}\bigg\{\sqrt{\frac{{\cal D}^2+(1-\Pi^2)R^2}
{1+R_z^2}}-2hR^2\bigg\}=0.
\end{equation}
Introducing the integration constant C, we find
\begin{eqnarray}\label{statsol}
R_z&=&\pm\frac{1}{2hR^2+C}\sqrt{{\cal D}^2+(1-\Pi^2)R^2-(2hR^2+C)^2} \nonumber\\
&=&\pm\frac{2h}{2hR^2+C}\sqrt{(R_+^2-R^2)(R^2-R_-^2)}.
\end{eqnarray}
Here $R_{\pm}$ are defined as
\begin{equation}
R_{\pm}^2=\frac{1}{8h^2}\bigg\{(1-\Pi^2-4hC)\pm\sqrt{(1-\Pi^2-4hC)^2+16h^2({\cal
D}^2-C^2)}\bigg\}.
\end{equation}
We also rewrite the BI field strength for further convenience as
\begin{equation}
E=\frac{{\cal D}}{\lambda (2hR^2+C)},\qquad B=\frac{\Pi
R^2}{\lambda (2hR^2+C)}.
\end{equation}
Though (\ref{statsol}) can always be integrated in terms of
elliptic integrals as done in \cite{Hyakutake4}, we have no need
to do it for general case.

In order to gain some insight we shall have some brief discussion
about the solution in the absence of RR fields. If we set $h=0$,
(\ref{statsol}) becomes $R_z=\pm\frac{1}{C}\sqrt
{(1-\Pi^2)R^2+{\cal D}^2-C^2}$. The equation for $R$ can be easily
integrated. The solution for the case $C={\cal D}$ is $R=\exp
{\big\{\pm \frac{\sqrt{1-\Pi^2}}{{\cal D}}(z -z_0)\big\}}$. This
solution leads a logarithmic bending of D2-brane. The electric
potential also admits a form of logarithm, while the magnetic
field is proportional to $R$. These are the characteristics of the
BIon solution. Actually It is just the BIon configuration
preserving $1/4$ supersymmetry found in \cite{Townsend1}. The
solution corresponding to $C^2>{\cal D}^2$ is
$R=\sqrt{\frac{C^2-{\cal D}^2}{1-\Pi^2}}\cosh {\big\{\pm
\frac{\sqrt{1-\Pi^2}}{C} (z -z_0)\big\}}$. This is electric
neutral catenoidal solution representing two D-branes joined by a
throat. The case $C^2<{\cal D}^2$ implies the solution
$R=\sqrt{\frac{{\cal D}^2-C^2}{1-\Pi^2}}\sinh{\big\{\pm
\frac{\sqrt{1-\Pi^2}}{C} (z -z_0)\big\}}$, which is a singular
deformation of the BIon solution as discussed in \cite{Gibbons}.

For $h$ nonzero, we note that the term in the square root of
(\ref{statsol}) should be larger than zero, so $R$ is restricted
between $R_{-}$ and $R_{+}$. To be more precisely $R$ can change
from $R_{-}$ to $R_{+}$ or from $R_{+}$ to $R_{-}$. General
solution extending in $z$ direction can be composed through
periodic continuation. If $R_-<0$, i.e. $C^2<{\cal D}^2$, there is
a singularity at finite $z$ as $R$ approach zero. As expected from
the case $h=0$, this represents a spike with finite length. Only
the case $R_-=0$ admits an infinitely long spike without
singularity. This is the main configuration we interested in
throughout this section.

In this case we can integrate (\ref{statsol}) easily to find
\begin{equation}\label{boa}
\sqrt{R_{+}^2 - R^2} +{{\cal D}\over 2h R_{+}} \ln{R_{+} +
\sqrt{R_{+}^2 - R^2}\over R} = |z - z_0|,
\end{equation}
with $R_{+}^2=\frac{1-\Pi^2-4h{\cal D}}{4h^2}$. In fact, the
solution (\ref{boa}) describes the BIon configuration. As $R$
approach $R_{+}$ the second term can be neglected. This leads to
$R^2+|z - z_0|^2=R_{+}^2$ which describes a sphere at $z=z_0$ with
radius $R_{+}$. As $R$ approach zero the second term is dominant
and $R\propto\exp{\big(-\frac{1}{L}(|z - z_0|-R_{+})\big)}$. This
is exactly the BIon solution satisfying the equation
$dz=\pm\frac{L}{2}\frac{dR^2}{R^2}$, with $L_s={{{\cal
D}}\over{2hR_{+}}}$ characterizing the length of the spike.
Furthermore the BI field strength can be rewritten as
\begin{equation}
F= {{\cal D}\over 2\lambda hR_{+} R}dt\wedge dR+ {\Pi R \over
2\lambda hR_{+}}dR\wedge d\theta.
\end{equation}
This suggest a radial Coulomb-like charge on the worldvolume, as
expected from the BIon solution. So the whole solution
representing a BIon configuration with a sphere. Similar
configuration was discovered in \cite{Emparan} with only the
electric field turned on, which is the sphalerons on top of a
potential barrier. But as we shall see later, when magnetic field
exist, this configuration would be stable.

The energy of the configuration (\ref{boa}) is given by
\begin{equation}\label{energy2}
H=2h T_2 V_3 + T_0 N\Pi+ 2\pi T_2 {\cal D}\int dz,
\end{equation}
where $V_3={4\over3}\pi R^3$ is the volume of the sphere. As shown
in \cite{Maldacena} the quantization condition on the ${\cal
D}$-flux is ${1\over 2\pi}\int d\theta {\cal D}=ng$, i.e., ${\cal
D}=ng$. By using this relation the last term in (\ref{energy2})
can be written as $nT_f\int dz$ which is just the energy of n
fundamental string. So one would observe that the energy exactly
split into three parts arising from the energy of RR flux,
D0-branes and fundamental strings. So we can identify this
configuration as the bound state of D2-brane with N D0-branes and
n fundamental strings dissolved in the worldvolume. Furthermore it
is noticed that the second term in (\ref{energy2}) is not the pure
energy of N D0-branes, but includes the contribution of ${\cal D}$
through $\Pi$. This can be explained by considering the
interaction of the fundamental strings and D0-branes. The N units
magnetic fluxes can be written as
\begin{equation}
N=\frac{\Pi}{h\lambda}R_+=\frac{\Pi\sqrt{\Lambda^2-\Pi^2}}{2h^2
\lambda},
\end{equation}
from which we can express $\Pi^2$ as a function of N
\begin{equation}
\Pi^2=\frac{1}{2}\bigg(\Lambda\pm \sqrt{\Lambda^2-4\Xi^2}\bigg).
\end{equation}
Here $\Lambda^2\equiv 1-4h{\cal D}$ and $\Xi\equiv 2h^2\lambda N$.
We consider the case with small BI field that $\Xi\ll 1$ and
$\Xi\ll\Lambda$, which is appropriate for our discussion below and
for later comparison with D0-brane description. The two possible
value of $\Pi^2$ can be approximated as
$\Pi^{2}_1=\Lambda^2-\frac{\Xi^2}{\Lambda^2}$ and
$\Pi^{2}_2=\frac{\Xi^2}{\Lambda^2}$. Inserting them back in
(\ref{energy2}) we can get the corresponding energy ( below we
have neglected the energy of the BI-string since it is always the
same in each case)
\begin{equation}\label{sphere}
H=\left\{
\begin{array}{ll}
\frac{NT_0}{\Lambda}(\Lambda^2-\frac{\Xi^2}{6\Lambda^2})& \qquad R_1=\frac{\Xi}{2h\Lambda}\\
\frac{1}{3}NT_0\frac{\Lambda^3}{\Xi}& \qquad
R_2=\frac{\Lambda}{2h}
\end{array}
\right.
\end{equation}

The situation with only magnetic or electric field had been
explored extensively. Now we shall investigate what will happen
when we turn on both fields. First with no magnetic fluxes, there
exist two configurations with $R_1=0$ and
$R_2=\frac{1}{2h}\sqrt{1-4h{\cal D}}$. They are just the stable
BI-string and the $S^2$ configuration with string ending as
discussed in \cite{Emparan}. As we turn on the magnetic fluxes, we
have two configuration with string ending on $S^2$, which is
different from the case with only electric field. One of them is
with a smaller sphere of which the radius is
$R_1=\frac{\Xi}{2h\Lambda}$, and the other is with a larger sphere
of which the radius is $R_2=\frac{\Lambda}{2h}$. It can be seen
from (\ref{sphere}) that the former have lower energy than the
latter under the condition of small BI field, i.e., the former is
more stable than the latter. Furthermore, one can see that under
the same circumstance the energy of the latter configuration is
independent of the number of magnetic flux $N$. Hence, the
configuration with a larger sphere corresponds to the $S^2$
configuration with string ending in the case of no magnetic field,
and it is unstable. As discussed previously now BI-string would
gain an infinite energy shift and become unstable. However,
because the magnetic flux can be interpreted as a 'dissolved'
D0-brane charge, turning on the magnetic field implies that some
D0-branes is added to the system. Due to the Myers effect, a stuck
of D0-branes is blown-up a sphere. This makes the BIon
configuration with string ending on the smaller sphere become
stable. On the other words, as $N$ being turned off the bludge
shrinks to zero size and the BI-string is recovered. Similarly, we
can also start from the case with only magnetic field, where there
is a meta-stable $S^2$ known as the dielectric sphere. We know
that in type IIA superstring theory there exists the BIon
configuration which represents a fundamental string ending on a
bound of a D2-brane and D0-branes \cite{Townsend1}. From the
viewpoint of the world volume theory on the D2-brane, the
fundamental string is expressed as a source of a Coulomb-like
electric field. Thus, as we turn on electric field, the
fundamental string is effectively attached to the spherical brane
configuration. Conclusively, the meta-stable configuration is the
spherical brane configuration with spikes.

\section{Fuzzy BIon from dielectric D0-branes }

\indent

In this section, we shall show that the configuration we have
found in the previous section also exist as a bound state of N
D0-branes. We start with the low energy effective action of N
D0-branes, which is the non-abelian extension of BI action
proposed in \cite{Myers1}, the leading nontrivial terms are
\begin{equation}
S\sim -NT_0V_1+\frac{1}{4}\lambda^2T_0 \int dt Tr (2D_0\phi^i
D_0\phi^i+[\phi^i,\phi^j]^2)-\frac{4}{3}ih\lambda^2\mu_0\epsilon_{ijk}
\int dt Tr (\phi^i\phi^j\phi^k),
\end{equation}
where $D_0\phi^i=\partial_0\phi^i+i[A_0,\phi^i]$ and
$i,j,k=1,2,3$. We have consistently set $\phi^4,\cdots,\phi^9$ to
zero, and the RR background $F_{0123}^{4}=-4h$ has been introduced into
the action through Chern-Simons term. It should be noticed that
this action is valid only when the commutators
$i\lambda[\phi^i,\phi^j]$ are small enough, or $h^2\lambda N\ll1$.
The corresponding Lagrangian can be written as
\begin{equation}\label{D0lag}
L=-NT_0+\lambda^2 T_0 Tr\bigg\{\frac{1}{2}D_0\phi^i D_0\phi^i +
\frac{1}{4}[\phi^i,\phi^j]^2 - i\alpha\epsilon_{ijk}
\phi^i\phi^j\phi^k \bigg\}.
\end{equation}
Here we define $\alpha=\frac{4}{3}h$ for further convenience. The
equation of motion for $\phi^i$ and Gauss law constraint derived
directly from the Lagrangian are
\begin{equation}\label{eom}
-\big[D_0,[D_0,\phi^i]\big]+\frac{3i\alpha}{2}\epsilon_{ijk}[\phi^j,\phi^k]+\big[[\phi^i,\phi^k],\phi^k\big]=0,\\
\qquad \big[\phi^i,[D_0,\phi^i]\big]=0.
\end{equation}
Following \cite{Hyakutake2}, for configurations with axial
symmetry, we choose the ansatz
\begin{eqnarray}
\phi^1_{mn}&=&\frac{1}{2}\rho_{m+1/2}\delta_{m+1,n}+\frac{1}{2}\rho_{m-1/2}\delta_{m,n+1}, \nonumber\\
\phi^2_{mn}&=&\frac{i}{2}\rho_{m+1/2}\delta_{m+1,n}-\frac{i}{2}\rho_{m-1/2}\delta_{m,n+1}, \\
\phi^3_{mn}&=&z_m\delta_{m,n}, \qquad A_{0mn}=a_m\delta_{m,n}
\nonumber.
\end{eqnarray}
Here we only consider the static configuration. Thus, substituting
the ansatz into (\ref{eom}), we get
\begin{eqnarray}\label{eeom}
(a_{m+1}-a_{m})^2-(z_{m+1}-z_m)^2+3\alpha(z_{m+1}-z_m)+\frac{1}{2}[\rho_{m-1/2}^2-2\rho_{m+1/2}^2+\rho_{m+3/2}^2]=0,\nonumber\\
-\frac{3\alpha}{2}(\rho_{m-1/2}^2-\rho_{m+1/2}^2)-[\rho_{m+1/2}^2(z_{m+1}-z_{m})-\rho_{m-1/2}^2(z_{m}-z_{m-1})]=0,\\
\rho_{m+1/2}^2(a_{m+1}-a_{m})-\rho_{m-1/2}^2(a_{m}-a_{m-1})=0
\nonumber.
\end{eqnarray}

To solve the above equations, we first choose the ansatz
$\rho_{m+1/2}=\frac{3}{2}\alpha\sqrt{(j+m+1)(j-m)}$, which is
reasonable since $\phi^{i}$ just become the fuzzy sphere solution
found in \cite{Myers1} when $z_m=\frac{3}{2}\alpha m$. Now we
are to find general $z_m$ and $a_m$ as a solution of (\ref{eeom}).
Substituting $\rho_{m+1/2}$ into the second equation of
(\ref{eeom}) yields
\begin{equation}\label{dz}
z_{m+1}-z_m=\frac{3}{2}\alpha+\frac{C}{\rho_{m+1/2}^2}
\end{equation}
Then inserting this back into the first equation in (\ref{eeom}),
one finds
\begin{equation}\label{da}
a_{m+1}-a_m=\pm\frac{C}{\rho_{m+1/2}^2}.
\end{equation}
It can be easily checked that this solution also satisfies the
third equation in (\ref{eeom}). Here C is an arbitrary constant.
For our convenience we can set $z_0=a_0=0$. From
(\ref{dz}) and (\ref{da}) one can find
\begin{eqnarray}\label{z}
z_m&=&\frac{3}{2}\alpha m + sgn(m)\sum_{k=0}^{|m|-1} \frac{
C}{\rho_{m+1/2}^2}\nonumber, \\
a_m&=&\pm sgn(m)\sum_{k=0}^{|m|-1} {C\over\rho_{m+1/2}^2}.
\end{eqnarray}
The algebra describing the fuzzy configuration can be written as
\begin{equation}\label{algebra}
[\phi^1,\phi^2]=i\frac{3}{2}\alpha(\phi^3\mp A_0),\\ \qquad
[\phi^2,\phi^3]=i\frac{3}{2}\alpha\phi^1\pm[\phi^2,A_0],\\ \qquad
[\phi^3,\phi^1]=i\frac{3}{2}\alpha\phi^2\pm[A_0,\phi^1].
\end{equation}
If we have redefinitions $\phi^1\equiv\tilde{\phi^1}$,
$\phi^2\equiv\tilde{\phi^2}$ and $\phi^3\equiv\tilde{\phi^3} \pm
A_0$, (\ref{algebra}) can be rewritten in terms of
$\tilde{\phi^i}$ as $[\tilde{\phi^i},\tilde{\phi^j}]
=i\frac{3}{2}\alpha\epsilon_{ijk}\tilde{\phi^k}$. This is
essentially the $SU(2)$ algebra with a deformation in the $z$
direction corresponding to two lumps protruding in the north and
south pole of fuzzy $S^2$. The shape of the lump is completely
determined by $A_0$. For the bound brane configuration including
the string, which we are just interested in,  $A_0$ can not be set
to zero. The reason is that $A_0$ is related to the string charge
as explained in \cite{Hyakutake3}.

When $C$ equals to zero in (\ref{z}), we have
$\phi^3\equiv\tilde{\phi^3}$. And it is obvious that $\phi^i$
satisfy the $SU(2)$ algebra representing a fuzzy sphere of radius
$h\lambda N$. Thus we recover the fuzzy $S^2$ blown-up due to
Myers effect. But the solution is more interesting with nonzero
$C$. The nonvanish contribution of the lump is described by
\begin{equation}
\Delta z_m=-\frac{1}{2}\frac{C}{4mh^2}\frac{\Delta
\rho_m^2}{\rho_{m+1/2}^2}.
\end{equation}
Here we use the identity $\Delta \rho_m^2=-8mh^2$. This is just
the regularized version of the differential equation that BIon
should satisfy. The coefficient $L_s\equiv \frac{C}{4mh^2}$ is
identified with the length of the spike we defined previously. In
the spike-dominant region $m\sim {N \over 2}$, the length is
approximate to $\lambda C\over\Xi$. It is easily seen that the
length is inverse proportional to the $\Xi$, which agree with
previous result implying the spike would be suppressed by the
magnetic flux. Furthermore, if we identify $\lambda C$ with $\cal
D$, the radius of the sphere and length of the spike are given by
$R=h\lambda N$ and $L_s={{\cal D}\over \Xi}$ respectively, which
is completely the same as the previous results for small $\cal D$

In order to analyze the stability of our solution, we should
calculate the correspondent energy. The Hamiltonian can be derived
from (\ref{D0lag}) as
\begin{equation}
H=NT_0+\lambda^2T_0 Tr
\bigg({1\over2}\partial_0\phi^i\partial_0\phi^i+
{1\over2}[A_0,\phi^i]^2-{1\over4}[\phi^i,\phi^j]^2+i\alpha\epsilon_{ijk}\phi^i\phi^j\phi^k\bigg).
\end{equation}
Imposing the static condition and using (\ref{eom}) to eliminate
the term with $A_0$ yields
\begin{equation}
H=NT_0-{1\over4}\lambda^2T_0Tr\bigg(i\alpha\epsilon_{ijk}\phi^i[\phi^j,\phi^k]-[\phi^i,\phi^j]^2\bigg).
\end{equation}
Substituting the solution (\ref{z}) into the Hamiltonian, we find
\begin{eqnarray}
H&=&H_0-{1\over2}\lambda^2T_0C\sum_{m=-j}^j {
\bigg(2\alpha-4h^2\alpha|m|\sum_{k=0}^{|m|-1}{1\over\rho_{k+1/2}^2}
+ {C\over\rho_{m+1/2}^2} \bigg)}\\ \nonumber
&=&H_0-{1\over2}\lambda^2T_0C\sum_{m=-j}^j{\Delta
z_m}-{1\over3}h\lambda^2T_0C.
\end{eqnarray}
Here $H_0$ is the energy of the usual fuzzy sphere and equals to
$NT_0-{2\over 3}NT_0(h^2\lambda N)^2(1-{1\over N^2})$. It is
interesting to observe that $\sum_{m=-j}^j{\Delta z_m}$
corresponds to the length of a string through the fuzzy sphere.
The contribution to energy due to this string is similar to the
that of fundamental string except that it is negative. If the
radius $h\lambda N$ is fixed, the last term is inverse
proportional to $N$ and vanish in the large $N$ limit. Thus the
total energy is smaller than $H_0$ for $C>0$, which means that our
solution with spikes has lower energy than the fuzzy $S^2$. This
implies that the dielectric sphere is not stable and can grow
spikes to lower its energy if we allow $A_0$ to be nonzero.
Conclusively, the BIon configuration given by us is more stable
than the configuration of the fuzzy sphere done by Myers
\cite{Myers1}.

So from above analysis we can see that our solution of the N
D0-brane action represent a BIon configuration with fuzzy $S^2$
structure, Which in the large N limit gives the same configuration
found in the previous section when we analyze the D2-brane
worldvolume action. This is what we have expected, and it also
supply a new evidence for the equivalence of the two pictures.

\section{Conclusions}

\indent

Based on the physical consideration, here we have investigated the
D2-brane in uniform RR background while we have incorporate both
electric and magnetic field in the world volume of the brane. Some
new static solutions of this system have been found. One among
them is geometrically described by a two-sphere with a spike,
charactered by the BIon of string ending on the spherical
D2-brane. In the language of BIon configuration, it is a BIon
configuration composed of the bound brane of a D2-brane, $n$
fundamental strings and $N$ D0-branes. By product, we also obtain
a tube solution. They can be understood as the blown-up objects of
the BI-string due to the Myers effect if we turn on the magnetic
field. It has been shown that these configurations are classically
stable. Due to the presence of the RR-flux, the configurations
found above are quantum mechanically unstable, they would tunnel
to configurations with larger radius. Further work is needed to
elucidate the process through the construction of bounce solution
and calculation of the decay rate.

In order to investigate how the dielectric D2-brane is affected in
the presence of electric field, we have also analyzed the a system
of N D0-branes in the same RR background. We give a solution
describing a BIons configuration with fuzzy $S^2$ structure which
is the same as the D2-brane picture. By the redefinition of the
matrix field, it is shown that algebra of the fuzzy configuration
is a deformed $SU(2)$ algebra which has nontrivial topology. By
comparing the energy of this BIon configuration with that of the
Myers' fuzzy $S^2$ configuration, we find that our configuration
is more stable than Myers'.

\section{Acknowledgments}

We would like to thank Sen Hu and J.X. Lu for helpful discussions.
Jing Shao also thanks Interdisciplinary Center for Theoretical
Study of USTC in Hefei for hospitality during his staying in the
Institute. The work was partly supported by the NNSF of China
(Grant No.90203003) and by the Foundation of Education Ministry of
China (Grant No.010335025).



\begin{thebibliography}{[99]}


\bibitem{Polchinski} J. Polchinski, {\it TASI Lectures on
D-branes,} hep-th/9611050

\bibitem{Leigh} R.G. Leigh, Mod.Phys.Lett. A {\bf 4}, (1989) 2767, hep-th/9911136

\bibitem{Emparan} R. Emparan, {\it Born-Infeld Strings Tunneling to D-branes,} Phys.Lett. B {\bf 423} (1998) 71, hep-th/9711106

\bibitem{Myers1} R.C. Myers, {\it Dielectric Branes,} JHEP {\bf 9912} (1999) 022, hep-th/9910053

\bibitem{Townsend1} D. Mateos and P.K. Townsend, {\it Supertubs,} Phys.Rev.Lett. {\bf 87} (2001) 011602, hep-th/0103030;
D. Mateos, S. Ng and P.K. Townsend, {\it Tachyons, Supertubes and
Brane/Anti-Brane Systems,} JHEP {\bf 0203} (2002) 016,
hep-th/0112054.

\bibitem{Gibbons} G.W. Gibbons, {\it Born-Infeld particles and Dirichlet p-branes,} Nucl.Phys. B {\bf 514} (1998) 603, hep-th/9709027

\bibitem{Maldacena} C.G. Callan and J.M. Maldacena, {\it Brane Dynamics From the Born-Infeld Action,} Nucl.Phys. B {\bf 513} (1998) 198, hep-th/9708147

\bibitem{Hyakutake1} Y. Hyakutake, {\it Fuzzy BIon,} hep-th/0305019

\bibitem{Hyakutake4} Y. Hyakutake, {\it Torus-like Dielectric D2-brane,} JHEP {\bf 0105} (2001) 013, hep-th/0103146;
D.K. Park, S. Tamaryan, Y.-G. Miao and H.J.W. Muller-Kirsten, {\it Tunneling of Born-Infeld Strings to D2-Branes,} Nucl.Phys. B {\bf 606} (2001) 84, hep-th/0011116

\bibitem{Hyakutake2} Y. Hyakutake, {\it Notes on the Construction of the D2-brane from Multiple D0-brane,} hep-th/0302190

\bibitem{Hyakutake3} Y. Hyakutake, {\it Expanded Strings in the Background of NS5-branes via a M2-brane,a D2-brane and D0-branes,} JHEP {\bf 0201} (2002) 021, hep-th/0112073




\end{thebibliography}
\end{document}